\documentclass[12pt,a4paper]{article}
\usepackage{cite}
\usepackage[dvips]{graphicx}
\usepackage{amssymb,amsmath}
\begin{document}
\pagestyle{headings}
\baselineskip=24pt
\def\figurename{Fig.}
\large
\title{\Large
METHOD FOR CONSTRUCTING RHEOLOGICAL MODELS OF INCOMPRESSIBLE MEDIA
UNDER FINITE DEFORMATIONS
\thanks{We gratefully acknowledge the support  of the Russian
Foundation for Fundamental Research and Perm Regional Department
of Science and Industry (grant 02-01-96404)}
}
\author{{\Large A. L. Svistkov and O. A. Putilova}\\[0.5em]
        Institute of Continuous Media Mechanics UB RAS,\\
        1, Academic Korolev Str., 614013, Perm, Russia\\
        E-mail: svistkov@icmm.ru}
\date{}
\maketitle
\thispagestyle{empty}
{\bf Introduction.} The viscoelastic behavior and viscous-flow
properties of incompressible materials undergoing finite deformations
can be simulated in terms of models of integral and differential type.
The integral models describe the history of medium deforma\-tion
using integral equations. However, these equations are difficult
to generalize to complicated problems such as, for instance,
simulation of mass exchange processes in dissipative media. 
Attempts to present damage accumulation and thixotropy
of elastomeric materials in terms of integral models add
complexity to these models and hamper their practical application.

The differential models describe the rheological properties
of materials in terms of tensor internal variables and, as a rule,
assign them the physical meaning of stresses
  \cite{Dafalias91, Govindjee92, Hausler95, Holzapfel96a,
  Holzapfel96b, Lion96, Simo87}
or strains
  \cite{Govindjee97, Haupt00, Lion97a, Lion97b, Lion98,
  Miehe00, Reese98}.
These models can be conveniently represented with the aid of symbolic
diagrams illustrating the mechanical behavior of the medium.
In the paper presented here a fairly simple method of
obtaining constitutive equations is described. The method has been
constructed using symbolic diagrams. The main advantage of the
method is rather simple physical meaning of mathema\-tical
expressions. The method has its origin in the ideas reviewed in
works
  \cite{Palmov80, Palmov97, Svistkov01}.

\begin{figure}[t]
\begin{center}
\includegraphics[width=12cm,height=9cm]{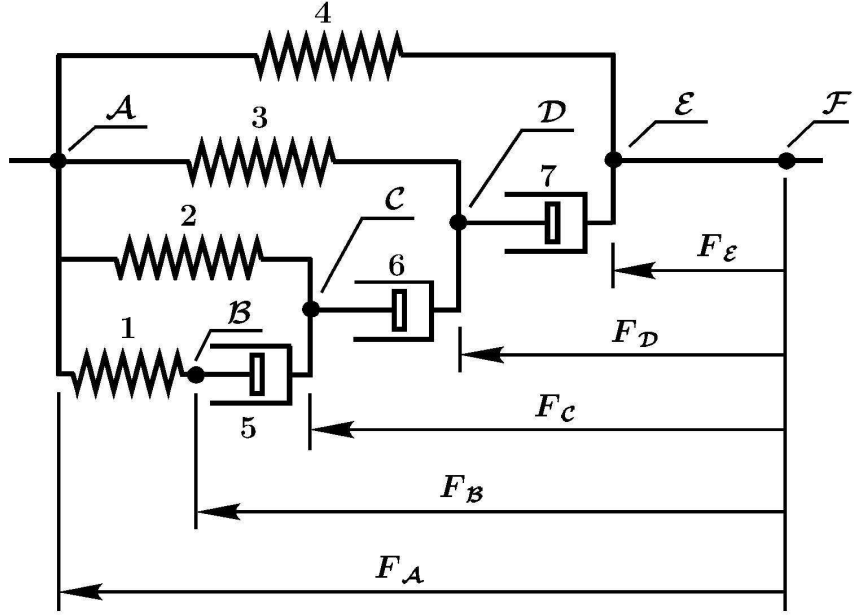}
\caption{\label{fig:kzaqerd}
Schematic representation of the model for viscoelastic material.
Arrows show linear mappings of the infinitesimal domain of the
material. Corresponding deformation gradients present information
about the mappings which couple reference and examined
points in the diagram and are used for introducing internal variables
(point coordinates of the diagram in nine-dimensional vector
space)}.
\end{center}
\end{figure}

{\bf Singular points in the symbolic diagram.} This work presents
the method allowing one to construct the mathematical model for
the medium whose mecha\-nical behavior is illustrated by a diagram
consisting of blocks of interrelated elastic and viscous elements.
Consider this method using the diagram shown in Fig.~\ref{fig:kzaqerd} as an
example. To each junction of elements from the symbolic diagram
illustrating the material behavior, a vector in 
nine-dimensional vector space corresponds. The space forms a set of
deformation gradient tensors. Between the points there are elastic and
viscous interrelations. Special role is assigned to three points: left,
right and reference.

{\it The left point in the scheme} is denoted by symbol
$\mathcal{A}$ (Fig.~\ref{fig:kzaqerd}) to which the deformation
gradient of the material corresponds. That is,
  $${\boldsymbol F}_\mathcal{A}={\boldsymbol F}=\,
  \frac{\partial{\boldsymbol x}}{\partial{\boldsymbol x}_0}\,,$$
where ${\boldsymbol x}={\boldsymbol x}(t,{\boldsymbol x}_0)$~is
the radius-vector of location of medium points at current time
$t$; ${\boldsymbol x}_0$~is the radius-vector of location of
medium points at reference time. It is evident that the rate of
deformation tensor ${\boldsymbol D}$ and the spin ${\boldsymbol
W}$, employed in the theory of nonlinear media, are connected with
point $\mathcal{A}$.

{\it The right point in the scheme} is denoted by symbol
$\mathcal{E}$ in Fig.~\ref{fig:kzaqerd}. Its peculiarity is that
the strain gradient ${\boldsymbol F}_\mathcal{E} $, corresponding
to point $\mathcal{E}$, is the rotation tensor ${\boldsymbol
R}_{\mathfrak{M}}$, which determines the rotation of the
infinitesimal bulk of the medium as an absolutely solid body and
is independent of the material deformation
  $${\boldsymbol F}_\mathcal{E}={\boldsymbol R}_{\mathfrak{M}}.$$
Hence, the rate of deformation tensor, calculated for the right
points of the system, is equal to zero
  $${\boldsymbol D}_\mathcal{E}=\,\frac{1}{2}\,\bigg(
  \stackrel{\;\centerdot}{\boldsymbol F}_\mathcal{E}
  \,{{\boldsymbol F}_\mathcal{E}}^{-1}+\,
  {{\boldsymbol F}_\mathcal{E}}^{-\rm T}\,
  \stackrel{\;\centerdot}{\boldsymbol F}_\mathcal{E}\!
  \vphantom{I}^{\rm T}\,\bigg)\,=0.$$

{\it The reference point} is denoted by symbol $\mathcal{F}$ in
Fig.~\ref{fig:kzaqerd}. In nine-dimensional space, the unit tensor
${\boldsymbol I}$ corresponds to this point
  $${\boldsymbol F}_\mathcal{F}={\boldsymbol I}.$$

\begin{figure}[t]
\begin{center}
\includegraphics[width=9cm,height=3cm]{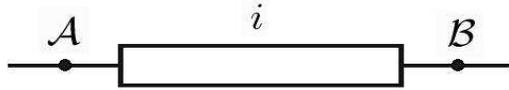}
\end{center}
\caption{\label{fig:kjwqerd} Diagram of the $i$th symbolic
element. In constructing of constitutive equations, deformation
gradients ${\boldsymbol F}_\mathcal{A}$, ${\boldsymbol
F}_\mathcal{B}$ and tensor measures of the rate of deformation of
${\boldsymbol D}_\mathcal{A}$, ${\boldsymbol D}_\mathcal{B}$}are
assigned to points $\mathcal{A}$ and $\mathcal{B}$.\end{figure}

{\bf Tensors used for establishing the strained state of the elements in
the symbolic diagram and the rate of change of this state.} Let us
examine some peculiarities of the mathematical description of the
strained state of the arbitrarily selected element of the system
(Fig.~\ref{fig:kjwqerd}). Assume that the element is numbered by $i$,
and its strained state is determined by comparing two deformation
gradients and two tensors of rate of deformation.

The element $i$ is connected with other elements at points
$\mathcal{A}$ and $\mathcal{B}$. Deformation gradients
${\boldsymbol F}_\mathcal{A}$ and ${\boldsymbol F}_\mathcal{B}$
correspond to points $\mathcal{A}$ and $\mathcal{B}$. Using these
gradients, the rate of deformation tensors ${\boldsymbol
D}_\mathcal{A}$, ${\boldsymbol D}_\mathcal{B}$ and spins
${\boldsymbol W}_\mathcal{A}$, ${\boldsymbol W}_\mathcal{B}$ are
determined as
  \begin{eqnarray}
  {\boldsymbol D}_\mathcal{A}&=&\frac{1}{2}\,\bigg(
  \stackrel{\
;\centerdot}{\boldsymbol F}_\mathcal{A}
  \,{{\boldsymbol F}_\mathcal{A}}^{-1}+\,
  {{\boldsymbol F}_\mathcal{A}}^{-\rm T}\,
  \stackrel{\;\centerdot}{\boldsymbol F}_\mathcal{A}\!
  \vphantom{I}^{\rm T}\,\bigg),
  \nonumber\\&&\nonumber\\
  {\boldsymbol D}_\mathcal{B}&=&\frac{1}{2}\,\bigg(
  \stackrel{\;\centerdot}{\boldsymbol F}_\mathcal{B}
  \,{{\boldsymbol F}_\mathcal{B}}^{-1}+\,
  {{\boldsymbol F}_\mathcal{B}}^{-\rm T}\,
  \stackrel{\;\centerdot}{\boldsymbol F}_\mathcal{B}\!
  \vphantom{I}^{\rm T}\,\bigg),
  \nonumber\end{eqnarray}
  \begin{eqnarray}
  {\boldsymbol W}_\mathcal{A}&=&\frac{1}{2}\,\bigg(
  \stackrel{\;\centerdot}{\boldsymbol F}_\mathcal{A}
  \,{{\boldsymbol F}_\mathcal{A}}^{-1}-\,
  {{\boldsymbol F}_\mathcal{A}}^{-\rm T}\,
  \stackrel{\;\centerdot}{\boldsymbol F}_\mathcal{A}\!
  \vphantom{I}^{\rm T}\,\bigg),
  \nonumber\\&&\nonumber\\
  {\boldsymbol W}_\mathcal{B}&=&\frac{1}{2}\,\bigg(
  \stackrel{\;\centerdot}{\boldsymbol F}_\mathcal{B}
  \,{{\boldsymbol F}_\mathcal{B}}^{-1}-\,
  {{\boldsymbol F}_\mathcal{B}}^{-\rm T}\,
  \stackrel{\;\centerdot}{\boldsymbol F}_\mathcal{B}\!
  \vphantom{I}^{\rm T}\,\bigg).
  \nonumber\end{eqnarray}\mbox{}

To interpret tensor values from physical standpoint, it is
reasonable to elaborate the appropriate mathematical model of the
medium starting from the following premises. The rate of deformation 
tensor of the $i$th element ${\boldsymbol D}_i$ can
be calculated {\it for viscous and elastic elements} from the
formula
  \begin{equation}
  {\boldsymbol D}_i\,=\,{\boldsymbol D}_\mathcal{A}-
  {\boldsymbol D}_\mathcal{B}.
  \label{eq:jnaewp}
  \end{equation}
Furthermore, in the elastic elements of the scheme the relation
between deformation gradients, corresponding to points 
$\mathcal{A}$ and $\mathcal{B}$, can be defined by the law of 
multiplicative decomposition
  \begin{equation}
  {\boldsymbol F}_\mathcal{A}\,=\,
  {\boldsymbol V}_i\,{\boldsymbol F}_\mathcal{B},
  \label{eq:jnaeeewp}
  \end{equation}
where the stretch tensor of the $i$th element ${\boldsymbol V}_i$
is found by three orthonormalized eigenvectors
$\boldsymbol{n}^i_1$, $\boldsymbol{n}^i_2$, $\boldsymbol{n}^i_k$
(defining the direction of principal tensor axes) and by
stretch ratios $\lambda^i_1$, $\lambda^i_2$, $\lambda^i_3$
  $${\boldsymbol V}_i=\,\sum^3_{k=1}\lambda^i_k\,
  \boldsymbol{n}^i_k\otimes\boldsymbol{n}^i_k.$$

It is easy to check that condition (\ref{eq:jnaeeewp}) determines
the relation between tensors ${\boldsymbol
D}_\mathcal{A}$, ${\boldsymbol W}_\mathcal{A}$ and ${\boldsymbol
D}_\mathcal{B}$, ${\boldsymbol W}_\mathcal{B}$ in elastic elements
  \begin{eqnarray}
  {\boldsymbol D}_\mathcal{B}&=&\frac{1}{2}\,
  {\boldsymbol V}_i^{-1}\,\Bigg(\,
  {\boldsymbol B}_i\,({\boldsymbol D}_\mathcal{A}+
  {{\boldsymbol W}_\mathcal{A}}^{\rm T})
  +({\boldsymbol D}_\mathcal{A}+
  {\boldsymbol W}_\mathcal{A})\,{\boldsymbol B}_i-\,
  \stackrel{\;\centerdot}{\boldsymbol B}_i\Bigg)\,
  {\boldsymbol V}_i^{-1},
  \label{eq:kmuhyg}\\
  &&\nonumber\\
  {\boldsymbol W}_\mathcal{B}&=&\frac{1}{2}\,
  {\boldsymbol V}_i^{-1}\,\Bigg(\,
  {\boldsymbol V}_i\stackrel{\;\centerdot}{\boldsymbol V}_i-
  \stackrel{\;\centerdot}{\boldsymbol V}_i{\boldsymbol V}_i +
  ({\boldsymbol D}_\mathcal{A}+
  {\boldsymbol W}_\mathcal{A})\,{\boldsymbol B}_i-
  \nonumber\\&&\nonumber\\&&\hspace{5em}
  -{\boldsymbol B}_i\,({\boldsymbol D}_\mathcal{A}+
  {{\boldsymbol W}_\mathcal{A}}^{\rm T})\Bigg)\,
  {\boldsymbol V}_i^{-1},
  \label{eq:kmuhygq}
  \end{eqnarray}
and the rate of time variation of the stretch ratio
  \begin{equation}
  \stackrel{\centerdot}{\lambda}\!\vphantom{l}^i_k=
  \lambda^i_k\,\boldsymbol{n}^i_k\otimes\boldsymbol{n}^i_k
  \cdot{\boldsymbol D}_i,
  \label{eq:wweon}
  \end{equation}
where the tensor ${\boldsymbol B}_i$ is the left Cauchy-Green tensor
of the elastic element
  $${\boldsymbol B}_i={{\boldsymbol V}_i}^2.$$

{\bf Coupling of elastic elements in the diagram.} In
the proposed method of constructing of constitutive equations
there is a condition which should be fulfilled when coupling 
the elements. The elastic elements must be connected 
by their left ends with other elastic elements or with the left point
in the symbolic diagram. This restriction is caused by the necessity 
to calculate by formula (\ref{eq:kmuhyg})
the time derivatives of tensors ${\boldsymbol B}_i$ for all 
elastic elements using for this tensors ${\boldsymbol D}_i$, 
${\boldsymbol D}$ and ${\boldsymbol W}$.
Calculation starts with elastic elements connected by their left ends
with the left-hand point in the diagram. At this step
we find not only the time derivatives of tensors ${\boldsymbol B}_i$,
but also the rate of deformation and spin tensors of the right
junction with the adjacent elements (tensors ${\boldsymbol
D}_\mathcal{B}$ and ${\boldsymbol W}_\mathcal{B}$ in
Fig.~\ref{fig:kjwqerd}). Calculations are performed using formulas
(\ref{eq:jnaewp}), (\ref{eq:kmuhyg}), (\ref{eq:kmuhygq}). 
Then we carry out calculations for elastic elements connected
on the right with the elements just examined, and so on.

{\bf Simulation of mechanical properties of medium
elements.} In isothermal case, as an expression for mass free-energy 
density $f$, we use the function of stretch ratios $\lambda^i_k$ of the elastic
elements of the medium
  \begin{equation}
  f=f(\lambda^1_1,\lambda^1_2,\lambda^1_3,...\,,\lambda^n_1,
  \lambda^n_2,\lambda^n_3),
  \label{eq:llljsas}
  \end{equation}
where $n$~is the number of elastic elements. It is assumed that
each elastic element satisfies the volume invariance condition
when subjected to deformation
  \begin{equation}
  \lambda^i_1\lambda^i_2\lambda^i_3=1.
  \label{eq:skdjfdq}
  \end{equation}
Therefore, the Cauchy stress tensor is calculated by the formula
from the nonlinear elastisity theory for incompressible materials
  \begin{equation}
  {\boldsymbol T}_i= p_i{\boldsymbol I}+
  \,\rho\sum^3_{k=1}\lambda^i_k
  \frac{\partial f}{\partial\lambda^i_k}\,\boldsymbol{n}^i_k
  \otimes\boldsymbol{n}^i_k,
  \label{eq:skdjoka}
  \end{equation}
where $\rho$~is the medium density, $p_i$~is an indefinite
parameter, and $i$~is the elastic element number.
     The characteristic features of viscous elements can be
simulated by ordinary state equations for viscous liquids
  \begin{equation}
  {\boldsymbol T}_j=p_j{\boldsymbol I}+2\,\eta_j{\boldsymbol D}_j
  \label{eq:skdjf}
  \end{equation}
and their incompressibility condition
  \begin{equation}
  {\boldsymbol I}\cdot{\boldsymbol D}_j=0,
  \label{eq:skdjp}
  \end{equation}
where ${\boldsymbol T}_j$~is the Cauchy tensor for viscous
elements, $\eta_j$~is the coefficient of shear viscosity, and $j$~is
the viscous element number.

{\bf Interrelations between stresses for coupled elements.} 
For determination of the closed system of constitutive equations,
it is necessary to establish the interrelations for stresses at 
diagram junctions. They must be formulated in
such a way that the sum of all forces acting on the arbitrarily
selected junction would be equal to zero. Since we study the
points in nine-dimensional vector space (space generates a set of
possible values of deformation gradient tensors), then we can use 
the Cauchy stress tensors of the diagram elements as the 
nine-dimensional force vectors.

Now let us formulate the rule. The sum of Cauchy stress tensors of
the elements coupled on the left is equal to the sum of Cauchy stress
tensors of the elements coupled on the right at any diagram junction.
It is assumed in this case that the forces, determined in nine-dimensional
vector space by the Cauchy stress tensor, act from 
left on the left point of the symbolic diagram and from 
right on its right point. It means that for the diagram given in
Fig.~\ref{fig:kzaqerd} the following equalities are valid:
  \begin{eqnarray}
  &&{\boldsymbol T}={\boldsymbol T}_1+
  {\boldsymbol T}_2+{\boldsymbol T}_3+{\boldsymbol T}_4
  \hspace{2em}(\mbox{at point }\mathcal{A}),
  \nonumber\\
  &&{\boldsymbol T}_1={\boldsymbol T}_5
  \hspace{9em}(\mbox{at point }\mathcal{B}),
  \nonumber\\
  &&{\boldsymbol T}_2+{\boldsymbol T}_5={\boldsymbol T}_6
  \hspace{6.5em}(\mbox{at point }\mathcal{C}),
  \nonumber\\
  &&{\boldsymbol T}_3+{\boldsymbol T}_6={\boldsymbol T}_7
  \hspace{6.5em}(\mbox{at point }\mathcal{D}).
  \nonumber \end{eqnarray}
The right point allows us to formulate the linearly dependent
equation
$${\boldsymbol T}_4+{\boldsymbol T}_7={\boldsymbol T}
  \hspace{6.5em}(\mbox{at point }\mathcal{E})$$
which gives no additional information.

{\bf Dissipation inequality.} In the isothermal case,
the thermodynamic constraint (second law of thermodynamics)
imposed on processes in the medium $${\boldsymbol
T}\cdot{\boldsymbol D}-
  \rho\stackrel{\centerdot}{ f}\,\ge 0$$
takes the form
$$\sum^n_{i=1}{\boldsymbol T}_i\cdot{\boldsymbol D}_i+
  \sum^{n+m}_{j=n+1}{\boldsymbol T}_j\cdot{\boldsymbol D}_j
  -\rho\stackrel{\centerdot}{ f}\,\ge 0,$$
where elastic and viscous elements are numbered by $i$ and $j$, 
respectively, $n$ is the number of elastic elements in the system,
and $m$ is the number of viscous elements in the system. Finally,
using the type of free energy (\ref{eq:llljsas}) and equlities
(\ref{eq:wweon}), (\ref{eq:skdjfdq})--(\ref{eq:skdjp}), the dissipation 
inequality can be written as
  $$\sum^{n+m}_{j=n+1}2\,\eta_j\,{\boldsymbol D}_j\cdot
  {\boldsymbol D}_j\ge 0.$$
This inequality is satisfied automatically provided that the shear
viscosity coefficients can not be negative.

\begin{figure}[t]
\begin{center}
\includegraphics[width=8cm,height=6cm]{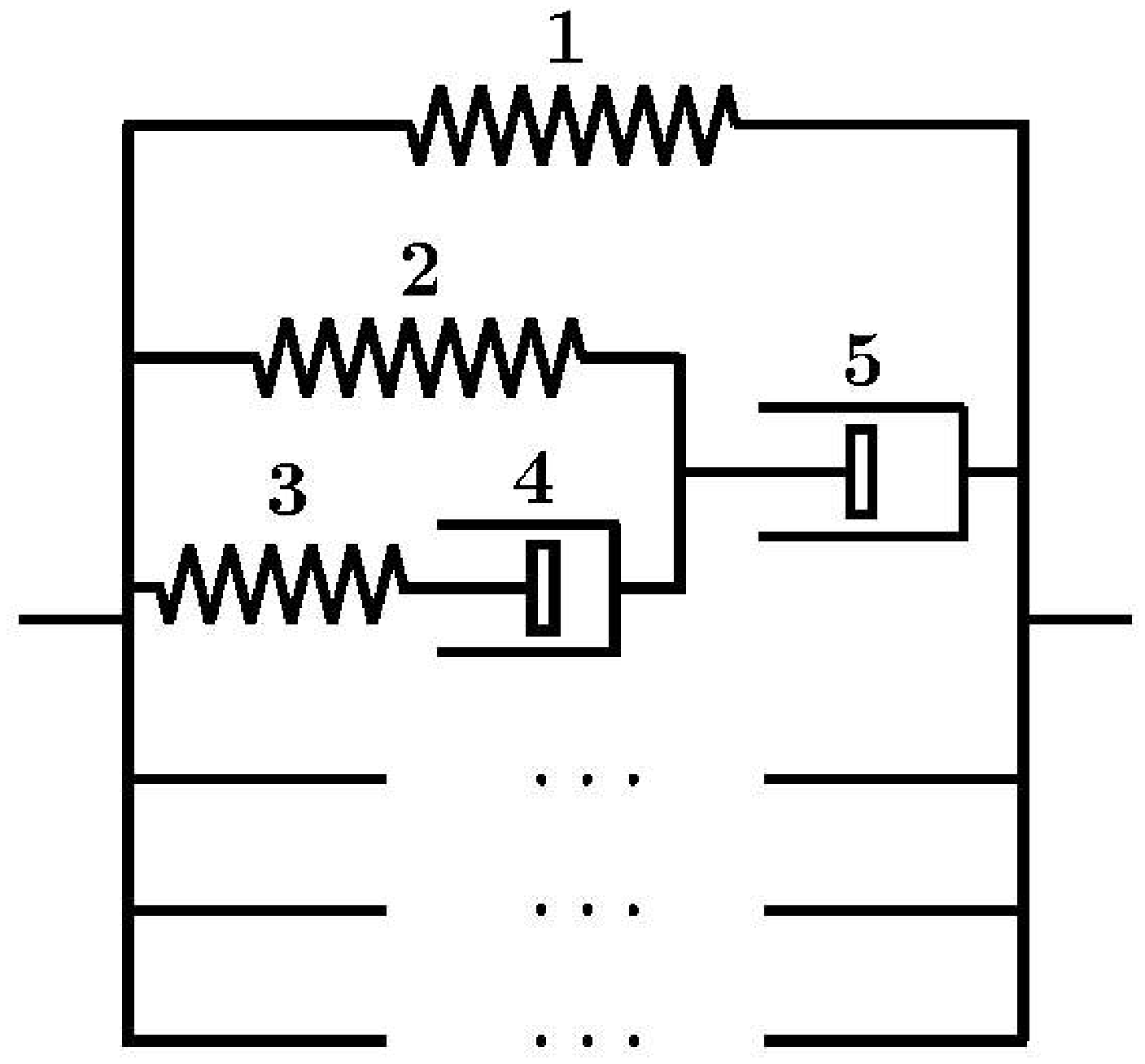}
\includegraphics[width=8cm,height=6cm]{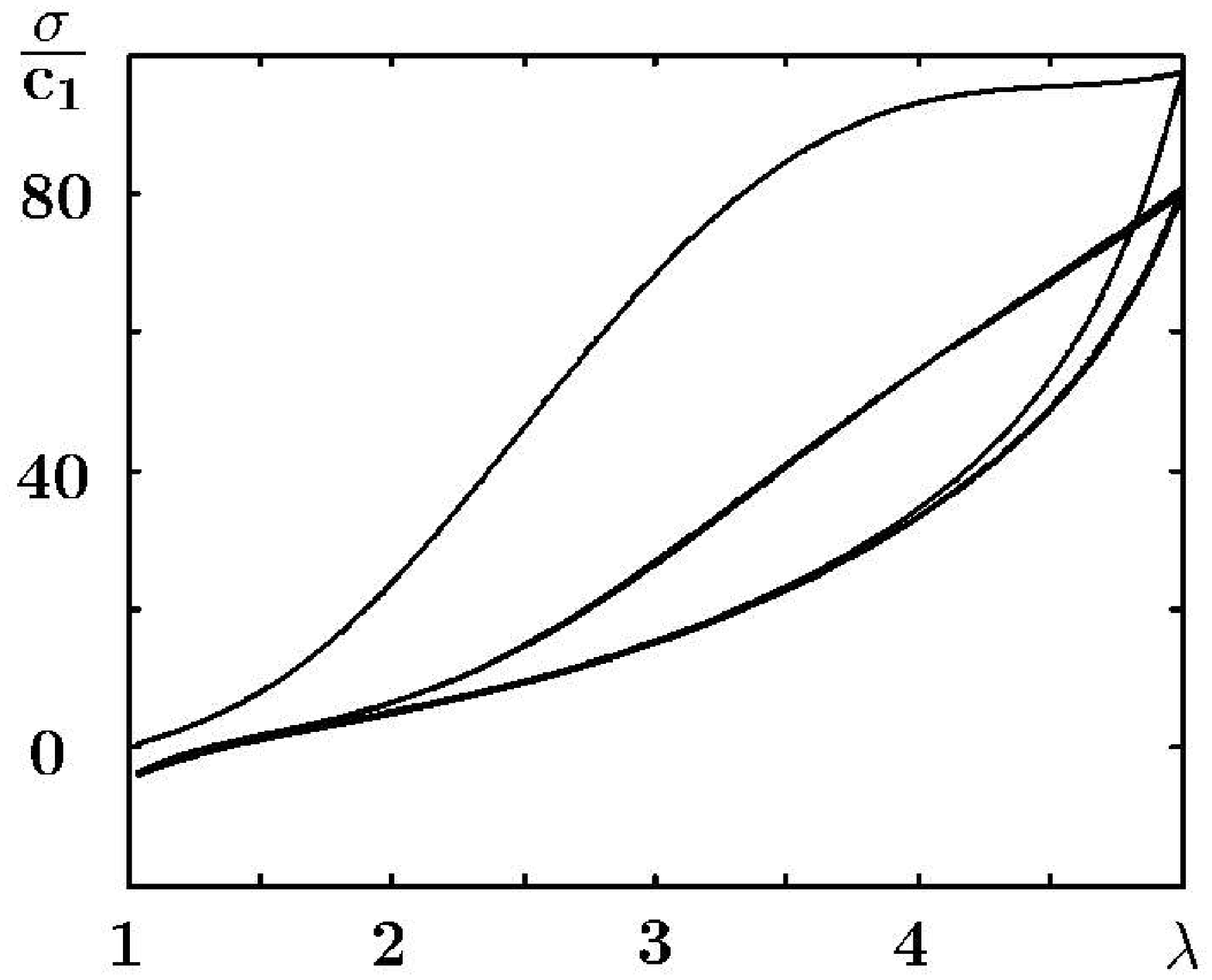}\\
\noindent
\mbox{}\hfil a \hfil\mbox{}\hfil b \hfil\mbox{}
\caption{\label{fig:qtttrd} Combination of elastic and viscous
elements (a) allows description of large hysteresis losses 
at large deformations and slightly changes the material behavior
at small deformations (b)}
\end{center}
\end{figure}

{\bf Example.} Consider the material whose behavior is illustrated
by a diagram in Fig.~\ref{fig:qtttrd}. Dots denote the blocks of
elements. We will not describe here the way of coupling elements
into these blocks. 

In modeling the viscoelastic properties of polymers, a proper
consideration of hysteresis losses is of particular importance. Usually 
the most challenging task is to provide the accurate description of large losses in the region of large values of stretch ratios and small losses in 
the region of small deformations of the material. In the considered example it is 
possible to describe such material behavior using the potential 
of free energy
  $$\rho f=-\sum^2_{i=1}\frac{c_i\,J^i_m}{6}\,
  \ln\,\bigg(1-\frac{I_1({\boldsymbol B}_i)-3}{J^i_m}\,\bigg)
  +c_3\,\bigg(I_1({\boldsymbol B}_1)-3\bigg)
  \bigg(I_1({\boldsymbol B}_3)-3\bigg),$$
where $c_1$, $c_2$, $c_3$, $J^1_m $, $J^2_m $~are material
constants, and $I_1({\boldsymbol B}_i)$~is the first invariant of
tensor ${\boldsymbol B}_i$. The first two terms in the potential
are known in the literature as Gent potentials formulated for
corresponding elastic elements \cite{Gent96}.

Two peculiarities of the material behavior defined by elements under 
consideration should be noted. During the first loading cycle there are
hysteresis significant losses and during the second and other cycles 
the losses are essentially less. 
However, such behavior of the material is viscoelastic by nature, and
not a manifestation of Mullins' softening effect.

Another distinguishing feature of the model is that the free
energy potential may involve the cross terms interrelating the
behavior of different elastic elements. In the example presented,
this is the last term. It means that the
deformation of the third element changes the elastic behavior of
the first, and the deformation of the first element specifies the
properties of the third. The physical meaning
of symbolic elements remains unchanged.

\def\refname{}

\end{document}